\documentclass{aastex631}

\graphicspath{{./}{figures/}}

\usepackage{graphicx}
\usepackage[caption=false]{subfig}

\begin{document}

\title{NOEMA Detection of Circumnuclear Molecular Gas in X-ray Weak Dual Active Galactic Nuclei: No Evidence for Heavy Obscuration}



\author[0000-0001-9062-8309]{Meicun Hou}
\affiliation{Kavli Institute for Astronomy and Astrophysics, Peking University, Beijing 100871, China}
\email{houmc@pku.edu.cn}

\author[0000-0003-0355-6437]{Zhiyuan Li}
\affiliation{School of Astronomy and Space Science, Nanjing University, Nanjing 210023, China}
\affiliation{Key Laboratory of Modern Astronomy and Astrophysics (Nanjing University), Ministry of Education, Nanjing 210023, China}

\author[0000-0003-0049-5210]{Xin Liu}
\affiliation{Department of Astronomy, University of Illinois at Urbana-Champaign, Urbana, IL 61801, USA}
\affiliation{National Center for Supercomputing Applications, University of Illinois at Urbana-Champaign, 605 East Springfield Avenue, Champaign, IL 61820, USA}

\author[0000-0002-7172-6306]{Zongnan Li}
\affiliation{National Astronomical Observatories, Chinese Academy of Sciences, 20A Datun Road, Chaoyang District, Beijing, China}

\author[0000-0001-8496-4162]{Ruancun Li}
\affiliation{Kavli Institute for Astronomy and Astrophysics, Peking University, Beijing 100871, China}
\affiliation{Department of Astronomy, School of Physics, Peking University, Beijing 100871, China}

\author[0000-0003-4956-5742]{Ran Wang}
\affiliation{Kavli Institute for Astronomy and Astrophysics, Peking University, Beijing 100871, China}
\affiliation{Department of Astronomy, School of Physics, Peking University, Beijing 100871, China}

\author[0000-0002-6593-8820]{Jing Wang}
\affiliation{Kavli Institute for Astronomy and Astrophysics, Peking University, Beijing 100871, China}

\author[0000-0001-6947-5846]{Luis C. Ho}
\affiliation{Kavli Institute for Astronomy and Astrophysics, Peking University, Beijing 100871, China}
\affiliation{Department of Astronomy, School of Physics, Peking University, Beijing 100871, China}

\begin{abstract}
Dual active galactic nuclei (AGN), which are the manifestation of two actively accreting supermassive black holes (SMBHs) hosted by a pair of merging galaxies, are a unique laboratory for studying the physics of SMBH feeding and feedback during an indispensable stage of galaxy evolution.
In this work, we present NOEMA CO(2-1) observations of seven kpc-scale dual-AGN candidates drawn from a recent {\it Chandra} survey of low-redshift, optically classified AGN pairs. These systems are selected because they show unexpectedly low 2--10 keV X-ray luminosities for their small physical separations signifying an intermediate-to-late stage of merger.  
Circumnuclear molecular gas traced by the CO(2-1) emission is significantly detected in 6 of the 7 pairs and 10 of the 14 nuclei, with an estimated mass ranging between $(0.2 - 21) \times10^9\rm~M_{\odot}$.
The primary nuclei, i.e., the ones with the higher stellar velocity dispersion, tend to have a higher molecular gas mass than the secondary.
Most CO-detected nuclei show a compact morphology, with a velocity field consistent with a kpc-scale rotating structure.
The inferred hydrogen column densities range between $5\times10^{21} - 2\times10^{23}\rm~cm^{-2}$, but mostly at a few times $10^{22}\rm~cm^{-2}$, in broad agreement with those derived from X-ray spectral analysis. 
Together with the relatively weak mid-infrared emission, the moderate column density argues against the prevalence of heavily obscured, intrinsically luminous AGNs in these seven systems, but favors a feedback scenario in which AGN activity triggered by a recent pericentric passage of the galaxy pair can expel circumnuclear gas and suppress further SMBH accretion.
\end{abstract}



\defcitealias{Hou2020}{H20}

\section{Introduction} \label{sec:intro}

Galaxy mergers, during which two galaxies exert strong gravitational force on each other and reshape the structure of their stellar and gaseous content, have long been recognized as a fundamental process in galaxy evolution \citep{Toomre1972,Barnes1992}. 
In particular, mergers tend to drive a gas inflow toward the center of either galaxy, or both, potentially triggering nuclear star formation \citep{Olson1990,Barton2000,Nikolic2004,Kewley2006}. 
In the meantime, a central supermassive black hole (SMBH), if it exists, can be fed by the same gas inflow and become an active galactic nucleus \citep[AGN;][]{Kauffmann2000,DiMatteo2005,Goulding2018}.
In principle, a galaxy merger can trigger a pair of AGNs, i.e., two simultaneously actively accreting SMBHs, also conventionally referred to as ``dual AGNs'' for pairs with a projected separation $r_{\rm p} \lesssim$ 10 kpc. 
Dual AGNs are valuable laboratories not only because they provide a crucial observational test for the theory and numerical simulations of galaxy mergers, but also because they are unique targets for studying the rich astrophysics involved in the feeding and feedback of SMBHs during an indispensable stage of galaxy evolution (see recent review by \citealp{DeRosa2019}).

Over the past decade, a number of systematic searches for dual-AGN candidates have been conducted, primarily in the optical band thanks to wide-field spectroscopic surveys such as the Sloan Digital Sky Survey (SDSS). 
In particular, the search for galactic nuclei with double-peaked narrow emission lines (e.g., [O III]$\lambda 5007$; \citealp{Comerford2009,Liu2010,Smith2010,Ge2012,LyuLiu2016}) aims at tight AGN pairs (typically 1--10 kpc in separation, but even less) that pertain to the late stage of mergers, while the search for spatially resolved AGN pairs showing optical emission-line characteristics of Seyferts or low-ionization nuclear emission-line regions (LINERs) covers a wider range of projected separation up to $r_{\rm p} \approx$ 100 kpc \citep{Liu2011}.
Confirmation of the AGN nature in these optically selected candidates, however, often requires follow-up observations in the X-ray and/or radio bands \citep{Comerford2011,Teng2012,Liu2013,Gabanyi2016,Brightman2018,DeRosa2018,Hou2019,Peng2022}, which are generally thought to trace immediate radiation from the SMBH and tend to be more immune to obscuration by circumnuclear cold gas. 
Infrared observations have also played an effective role in revealing dual AGNs, especially in gas-rich merging systems \citep{Satyapal2014,Satyapal2017,Weston2017,Ellison2017}.
Alternatively, one starts with a hard ($\gtrsim$ few keV) X-ray selected AGN and tries to associate it with another AGN in a companion galaxy, if present \citep{Koss2012}. Or, one tries to identify dual AGNs directly from high-resolution radio images \citep{Fu2015a,Fu2015b}.
These approaches have achieved varied degrees of success, revealing a growing number of dual AGNs at low redshift ($z \lesssim 0.5$). 

Despite the continued observational effort, a consensus on the occupation rate of AGNs in mergers, and more generally in interacting galaxies, is still absent. 
This is tied to the challenge of detecting low-luminosity AGNs (i.e., bolometric luminosity $L_{\rm bol} \lesssim 10^{43}~{\rm erg~s^{-1}}$), which are the manifestation of SMBHs fed at relatively low rates.
Because less fuel is required, secular processes such as instabilities driven by bars and/or minor mergers, may be sufficient to trigger low-luminosity AGNs \citep{Ho2008,Ho2009,Hopkins2014,Menci2014}. 
This leaves the causality between galaxy interactions and AGN triggering an open question. 
Indeed, some observations found an excess of close neighbors in AGN host galaxies or a higher fraction of AGNs in interacting galaxies than in isolated galaxies \citep{Koss2010, Ellison2011,Silverman2011,Liu2012,Goulding2018}, whereas others detected no significant difference \citep{Ellison2008, Darg2010, Villforth2014, Villforth2017}. 
Recently, \citet{Hou2020} presented the first systematic X-ray study of optically selected AGN pairs at low redshift (median redshift $\bar{z} \sim 0.1$), based on a homogeneous sample of $\sim 10^3$ SDSS AGN pairs identified by \citet{Liu2011}. 
Utilizing archival {\it Chandra} observations, \citet{Hou2020} were able to detect or place significant constraint on the nuclear X-ray emission in a subset of 67 pairs, down to a limiting 2--10 keV X-ray luminosity $L_{\rm X} \sim 10^{40}\rm~erg~s^{-1}$. 
Moreover, interesting trends are revealed in the mean AGN X-ray luminosity as a function of projected separation:
First, $L_{\rm X}$ increases with decreasing $r_p$ when $r_p \gtrsim 15$ kpc, suggesting enhanced SMBH accretion even at the early stage of mergers. Second, $L_{\rm X}$ decreases with decreasing $r_p$ at $r_p \lesssim 15$ kpc, when $r_p \lesssim 5$ kpc falling to a mean value below that of single AGNs whose host galaxy properties have been matched to the AGN pair sample.
The latter trend is particularly surprising and not predicted by existing numerical simulations of galaxy mergers \citep[e.g.,][]{Capelo2015,Capelo2017,Yang2019}. 
\citet{Hou2020} considered two plausible explanations:
(i) At the intermediate-to-late stage of the merger, gas inflows have led to a central concentration of cold gas that heavily obscures even the hard (2--10 keV) X-rays; (ii) AGN feedback triggered by recent pericentric passage of the galaxy pair can expel gas from the central region and suppress subsequent SMBH accretion.

The first scenario can be directly tested by the detection (or non-detection) of circumnuclear molecular gas with an equivalent hydrogen column density as high as $10^{23}-10^{24} \rm~cm^{-2}$, required to block keV X-rays from the embedded AGN. 
To shed light on the cause of the apparently low nuclear X-ray luminosities in the merging pairs with the smallest projected separations ($r_p \lesssim 5$ kpc), 
we have conducted high-resolution CO observations using the IRAM NOrthern Extended Millimeter Array (NOEMA), aiming to detect circumnuclear molecular gas in a pilot sample drawn from \citet{Hou2020}.
Most previous studies of molecular gas in interaction galaxies and mergers have focused on gas-rich, (ultra-)luminous infrared galaxies \citep[LIRGs and ULIRGs; e.g.,][]{Evans2002,Sakamoto2008,Sakamoto2014,Feruglio2015,Sliwa2017,Herrera-Camus2020,Tan2021}, with specific interest in studying molecular outflows.
There have also been studies of molecular gas in normal interacting galaxies, which focus on the statistical relation between molecular gas and star forming activities \citep{Kaneko2013,Kaneko2017,Michiyama2016,Violino2018,Shangguan2019,Thorp2022}.
Relatively little attention has been paid to the relation between circumnuclear molecular gas and AGN activity in optically (via [O III]) and X-ray selected, kpc-scale mergers such as those studied here.


This paper is organized as follows. Section \ref{sec:sample} describes the target selection, NOEMA CO(2-1) observations and data reduction. The detection of CO(2-1) emission and the inferred properties of circumnuclear molecular gas in the dual-AGNs are presented in Section~\ref{sec:analysis}.  
Main conclusions and implications of this study are addressed in Section \ref{sec:discuss}.
Throughout this paper, we assume a concordance cosmology with $\Omega_m = 0.3$, $\Omega_{\Lambda} = 0.7$, and $H_{0}=70$ km s$^{-1}$ Mpc$^{-1}$.
Errors are quoted at 1$\sigma$ confidence level, unless otherwise stated.

\section{Sample Selection and Data Reduction} \label{sec:sample}

\subsection{Target Selection}

Our targets of seven dual-AGNs were drawn from the sample of \cite{Hou2020} (hereafter H20), which consists of 67 candidate AGN pairs with archival {\it Chandra} observations. 
The \citetalias{Hou2020} sample itself was drawn from a parent sample of 1286 optical spectroscopic AGN pairs selected from SDSS DR7 \citep{Liu2011}, which are dominated by Type 2 AGNs whose optical narrow emission-line ratios are characteristic of Seyferts, LINERs, and/or AGN-[H II] composites.
All those AGN pairs have projected physical separations $r_{\rm p} < 100$ kpc and line-of-sight velocity offsets $<$ 600 km s$^{-1}$.
Here for the pilot study, we selected the most closely separated dual-AGN candidates, which include seven 
systems with $r_{\rm p} < 4.5$ kpc.
In the X-ray (0.5--8 keV) band, 4 systems have both nuclei detected and 3 have one nucleus detected; when only the hard X-ray band (i.e., 2--10 keV) is considered, 5 of the 14 nuclei are detected.  
The 2--10 keV X-ray luminosities (or upper limits) of the 14 nuclei range between $4.6\times10^{40} - 1.9 \times10^{42}\rm~erg~s^{-1}$, and their mean luminosity is systematically lower than that of the more separated (i.e., $r_{\rm p} \gtrsim 5$ kpc) nuclei in the \citetalias{Hou2020} sample. 
The basic information of the 14 nuclei as well as their host galaxies are listed in Table~\ref{table:info}. 
We use `P' (`S') to denote the primary (secondary) nucleus in a pair, which has the higher (lower) stellar velocity dispersion measured from the SDSS single-fibre spectra\footnote{While stellar velocity dispersion is provided by the MPA-JHU catalog (\url{https://wwwmpa.mpa-garching.mpg.de/SDSS/DR7/}), here we have employed the penalized pixel-fitting (pPXF) algorithm \citep{Cappellari2004} to fit the SDSS spectra to extract the stellar kinematics. The best-fit stellar line-of-sight velocity and velocity dispersion are in good agreement with the MPA-JHU values in most cases.}.  

\begin{deluxetable*}{lcccccccc}
\tabletypesize{\footnotesize}
\tablecaption{Basic Information of Seven Dual-AGNs
\label{table:info}}
\tablehead{
\colhead{Full Name} & 
\colhead{Redshift} & 
\colhead{$\Delta$V} &
\colhead{$\Delta\theta$} &
\colhead{$r_{\rm p}$} & 
\colhead{$\sigma_{\ast}$} & 
\colhead{log $L_{\rm 2-10}$} &
\colhead{log $L_{\rm X}/L_{\rm Edd}$} &
\colhead{$W1-W2$}
}
\colnumbers
\startdata
J002208.69+002200.5 (S) &  0.0710  &  93 &  3.1  &  4.2  & $199.4\pm 6.4$ &  $<$ 40.85 (1)  & $< -$5.36 &  $ 0.03 \pm 0.03$ \\
J002208.83+002202.8 (P) &  0.0707  &  93 &  3.1  &  4.2  & $231.2\pm 5.2$ &  $<$ 40.91 (1)  & $< -$5.57 &  $ 0.03 \pm 0.03$ \\
\hline
J085837.53+182221.6 (S) &  0.0587  &  58 &  2.9  &  3.3  & $130.2\pm10.3$ &  $<$ 40.83 (1)  & $< -$4.59 &  $ 0.26 \pm 0.03$ \\
J085837.68+182223.4 (P) &  0.0589  &  58 &  2.9  &  3.3  & $186.5\pm 8.0$ &  $<$ 40.97 (1)  & $< -$5.11 &  $ 0.26 \pm 0.03$ \\
\hline
J102700.40+174901.0 (P) &  0.0665  &  28 &  2.4  &  3.0  & $156.2\pm 5.3$ &  $<$ 40.67 (0)  & $< -$5.09 &  $ 0.17 \pm 0.05$ \\
J102700.56+174900.3 (S) &  0.0666  &  28 &  2.4  &  3.0  & $133.2\pm 7.3$ &  $40.81^{+0.13}_{-0.15}$ (1)  & $-4.66^{+0.06}_{-0.15}$ &  $ 0.17 \pm 0.05$ \\
\hline
J105842.44+314457.6 (S) &  0.0728  & 172 &  2.9  &  4.1  & $108.8\pm16.6$ &  $<$ 41.28 (1)  & $< -$3.81 &  $ 0.46 \pm 0.04$ \\
J105842.58+314459.8 (P) &  0.0723  & 172 &  2.9  &  4.1  & $158.8\pm14.7$ &  $42.29^{+0.05}_{-0.05}$ (1)  & $-3.49^{+0.05}_{-0.05}$ &  $ 0.46 \pm 0.04$ \\
\hline
J133031.75-003611.9 (S) &  0.0542  &  23 &  4.2  &  4.4  & $ 85.8\pm12.2$ &  $<$ 41.02 (0)  & $< -$3.64 &  $ 0.60 \pm 0.03$ \\
J133032.00-003613.5 (P) &  0.0542  &  23 &  4.2  &  4.4  & $142.0\pm10.6$ &  $40.96^{+0.15}_{-0.19}$ (1)  & $-4.62^{+0.15}_{-0.19}$ &  $ 0.60 \pm 0.03$ \\
\hline
J154403.45+044607.5 (S) &  0.0420  & 113 &  4.1  &  3.4  & $124.4\pm13.4$ &  $<$ 40.78 (1)  & $< -$4.56 &  $ -0.11 \pm 0.23$ \\
J154403.67+044610.1 (P) &  0.0416  & 113 &  4.1  &  3.4  & $200.2\pm 8.7$ &  $41.55^{+0.06}_{-0.07}$ (1)  & $-4.67^{+0.06}_{-0.07}$ &  $ -0.11 \pm 0.23$ \\
\hline
J220634.97+000327.6 (S) &  0.0466  & 130 &  4.7  &  4.3  & $ 74.8\pm 9.0$ &  $<$ 40.86 (0)  & $< -$3.54 &  $ 0.14 \pm 0.03$ \\
J220635.08+000323.2 (P) &  0.0461  & 130 &  4.7  &  4.3  & $172.9\pm 8.6$ &  $41.75^{+0.05}_{-0.06}$ (1)  & $-4.19^{+0.05}_{-0.06}$ &  $ 0.14 \pm 0.03$ \\
\enddata
\tablecomments{
(1) SDSS names with J2000 coordinates given in the form of ``hhmmss.ss+ddmmss.s'', the targets are listed in the order of RA; `P' and `S' denote the primary and secondary nuclei according to the stellar velocity dispersion;
(2) spectroscopic redshift from the SDSS DR7, based on the fit to the stellar continuum; (3) velocity offset of the two nuclei in each pair, in units of km s$^{-1}$; (4)-(5) projected angular and physical separation of the two nuclei in each pair, in units of arcsecond and kpc, respectively; 
(6) stellar velocity dispersion, in units of km s$^{-1}$, derived with pPXF \citep{Cappellari2004};
(7) absorption corrected luminosities of nucleus in the 2--10 keV band with a presumed absorption column density of $10^{22}\rm~cm^{-2}$ and a power-law photon-index of 1.7, in units of ${\rm erg~s^{-1}}$, from \citetalias{Hou2020}; 1 and 0 in parentheses represent detection and non-detection in X-rays in the 0.5--8 keV band, respectively;
(8) X-ray Eddington ratio; 
(9) WISE $W1$(3.4 $\mu$m)$-W2$(4.6 $\mu$m) color, from \citet{Wright2010}, in units of mag.
}
\end{deluxetable*}

\subsection{NOEMA Observations and Data Reduction}\label{subsec:datareduction}

We observed the targets with NOEMA between October 2020 and January 2021 using the single-field observing mode under C-configuration (project: S20BJ; PI: M. Hou). A total of 9--11 antennae participated in the observations, depending on the target.
The observations aim to probe the circumnuclear molecular gas via the CO(2-1) line at a rest-frame frequency of 230.54 GHz. 
Compared to CO(1-0), CO(2-1) is more sensitive to the high column densities of molecular gas expected in the dual-AGNs and allows for a higher angular resolution of $\sim 0.9\arcsec$, which corresponds to a typical linear scale of 1 kpc at the distance of our targets and is sufficient to distinguish the two nuclei.
The exposure time for each target ranged from 0.4 hr to 4.5 hr. 
We used the PolyFiX correlator in Band 3. 
The tuning frequency was set according to the mean redshift of each target pair.
The flux calibrators included LKHA101, MWC349 and 1055+018, which have a systematic uncertainty of $\lesssim$ 20\%.

The data reduction and calibration were performed using the CLIC package of GILDAS (versions oct21a)\footnote{\url{https://www.iram.fr/IRAMFR/GILDAS}} following the standard procedure. The image cleaning and analysis were done with the MAPPING package of GILDAS. 
We binned the frequency channels by a factor of 5 to achieve an effective velocity resolution of $\sim 14~\rm{km~s^{-1}}$, which is optimal for the signal-to-noise ratio (S/N) of the line emission given the expected velocity dispersion of the circumnuclear gas.
We adopted the natural weighting, HOGBOM algorithm and cleaned the image down to the one time of the noise level determined by the initial clean version of the image with the central source region masked.
Basic information of the CO observations are reported in Table~\ref{table:CO}.

\begin{figure*}\centering
\subfloat{\includegraphics[width=0.92\textwidth,angle=0]{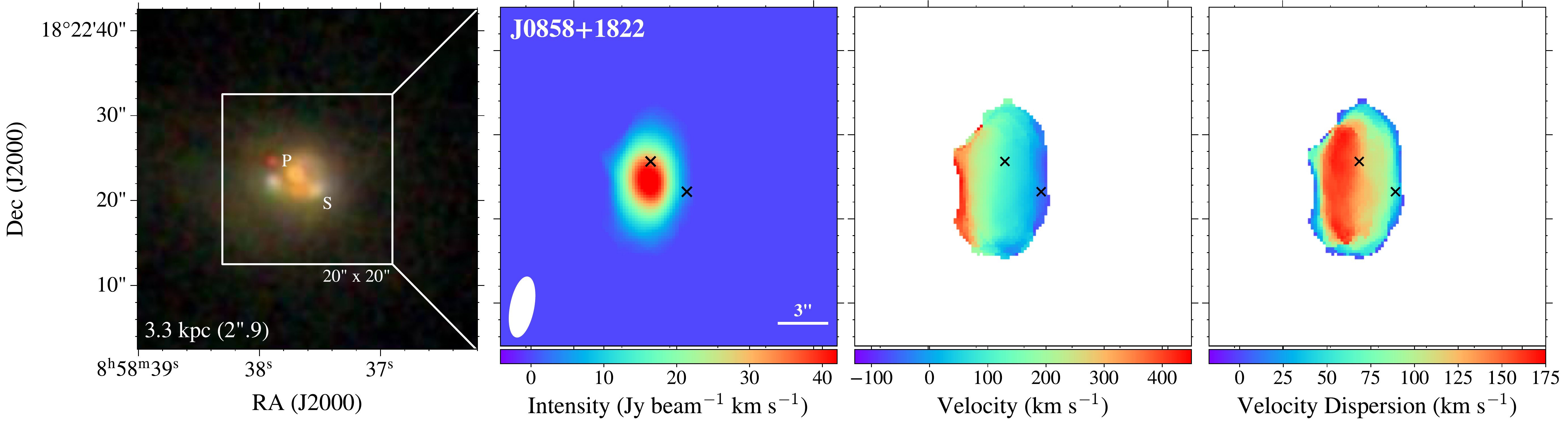}}
\vspace{-0.35cm}
\qquad 
\subfloat{\includegraphics[width=0.92\textwidth,angle=0]{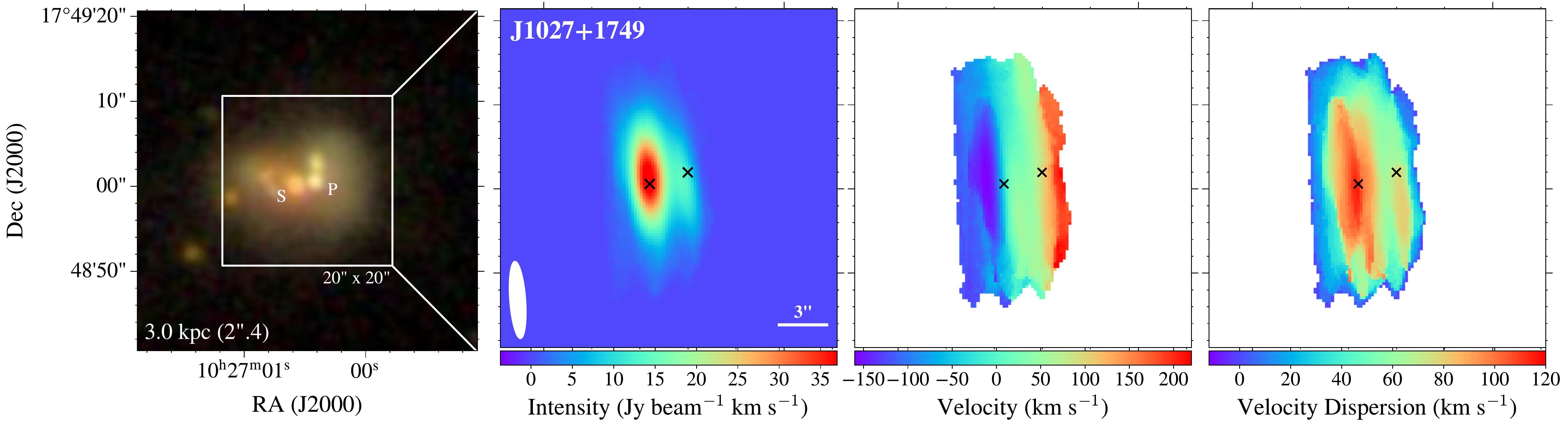}} 
\vspace{-0.35cm}
\qquad 
\subfloat{\includegraphics[width=0.92\textwidth,angle=0]{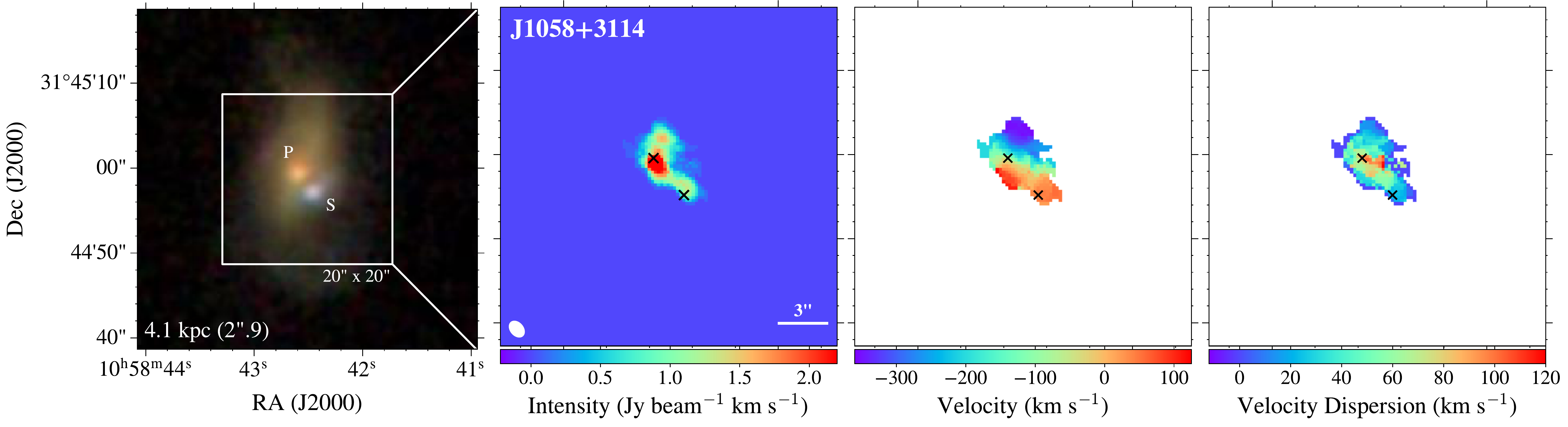}}
\caption{Six dual-AGN systems with at least one nucleus detected in CO(2-1) emission. The SDSS $gri$-color composite image is shown in left panel, and the NOEMA CO(2-1) moment 0, 1 and 2 maps (intensity, velocity, and velocity dispersion maps) are shown in the right three panels with a size of $20'' \times 20''$. North is up and east is to the left. The pairs are ordered in increasing RA. For each pair, the projected physical separation is labeled with angular distance in parenthesis. `P' and `S' denote the primary and secondary nuclei according to the stellar velocity dispersion. Crosses in the moment maps mark positions of the optical nuclei. The synthesized beam is showed in the lower-left corner of moment 0 map.
}
\label{fig:imagepair}
\end{figure*}

\begin{figure*}
  \ContinuedFloat 
\centering
\subfloat{\includegraphics[width=0.92\textwidth,angle=0]{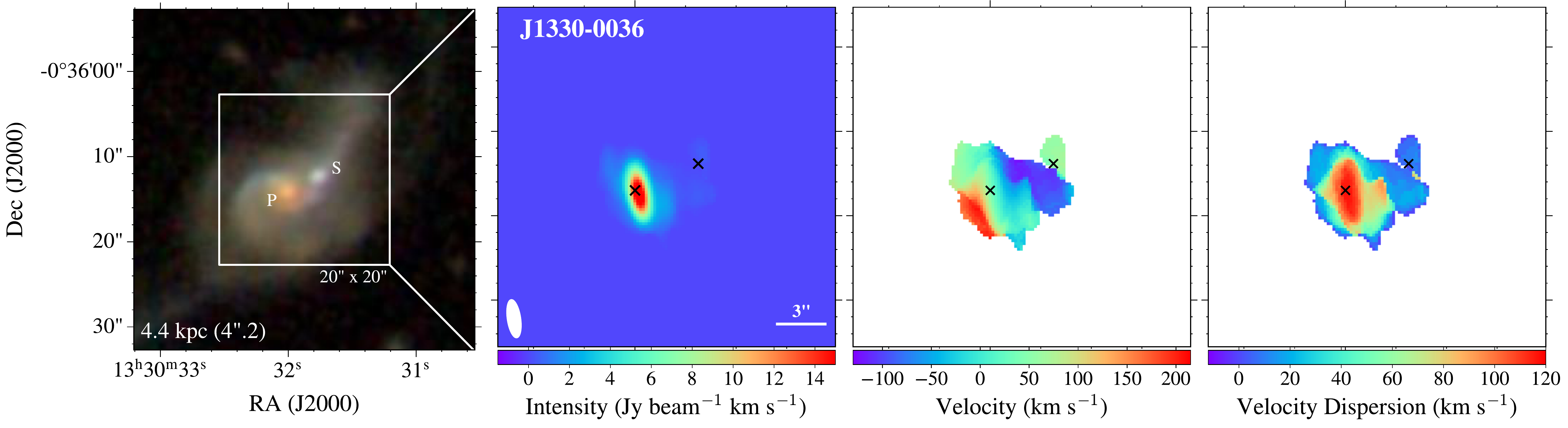}}
\vspace{-0.35cm}
\qquad 
\subfloat{\includegraphics[width=0.92\textwidth,angle=0]{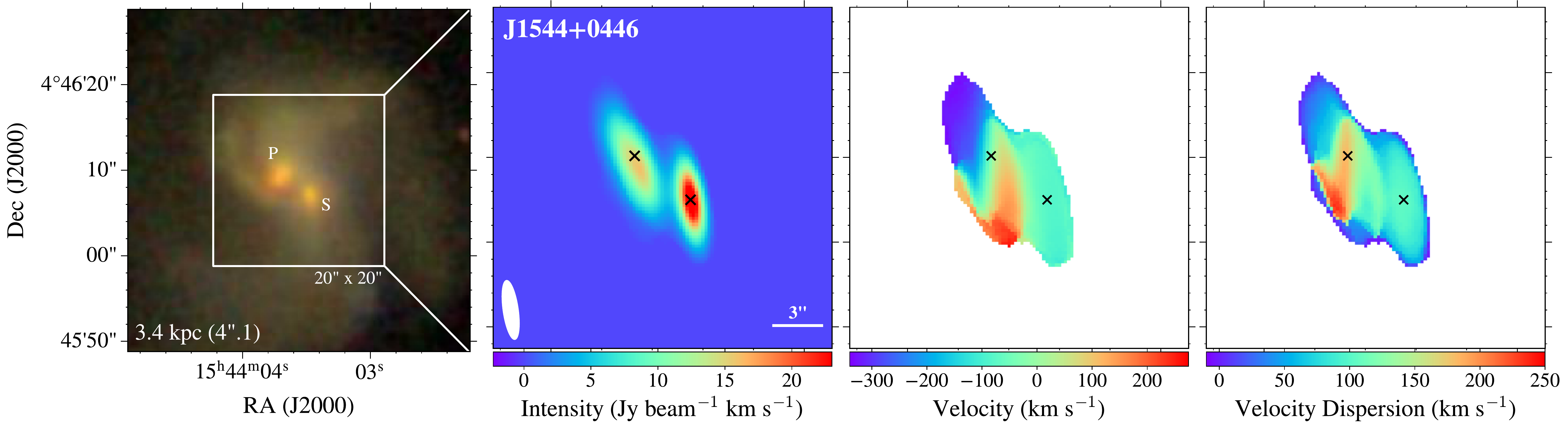}}
\vspace{-0.35cm}
\qquad 
\subfloat{\includegraphics[width=0.92\textwidth,angle=0]{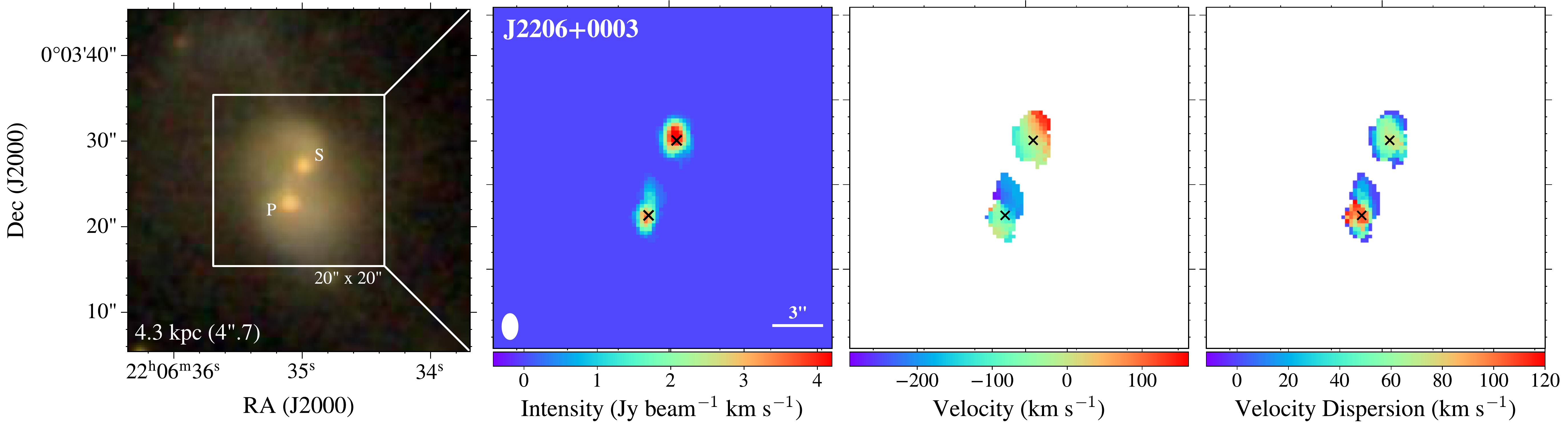}}
\caption{--continued}
\end{figure*}

\begin{figure*}\centering
\includegraphics[width=0.95\textwidth,angle=0]{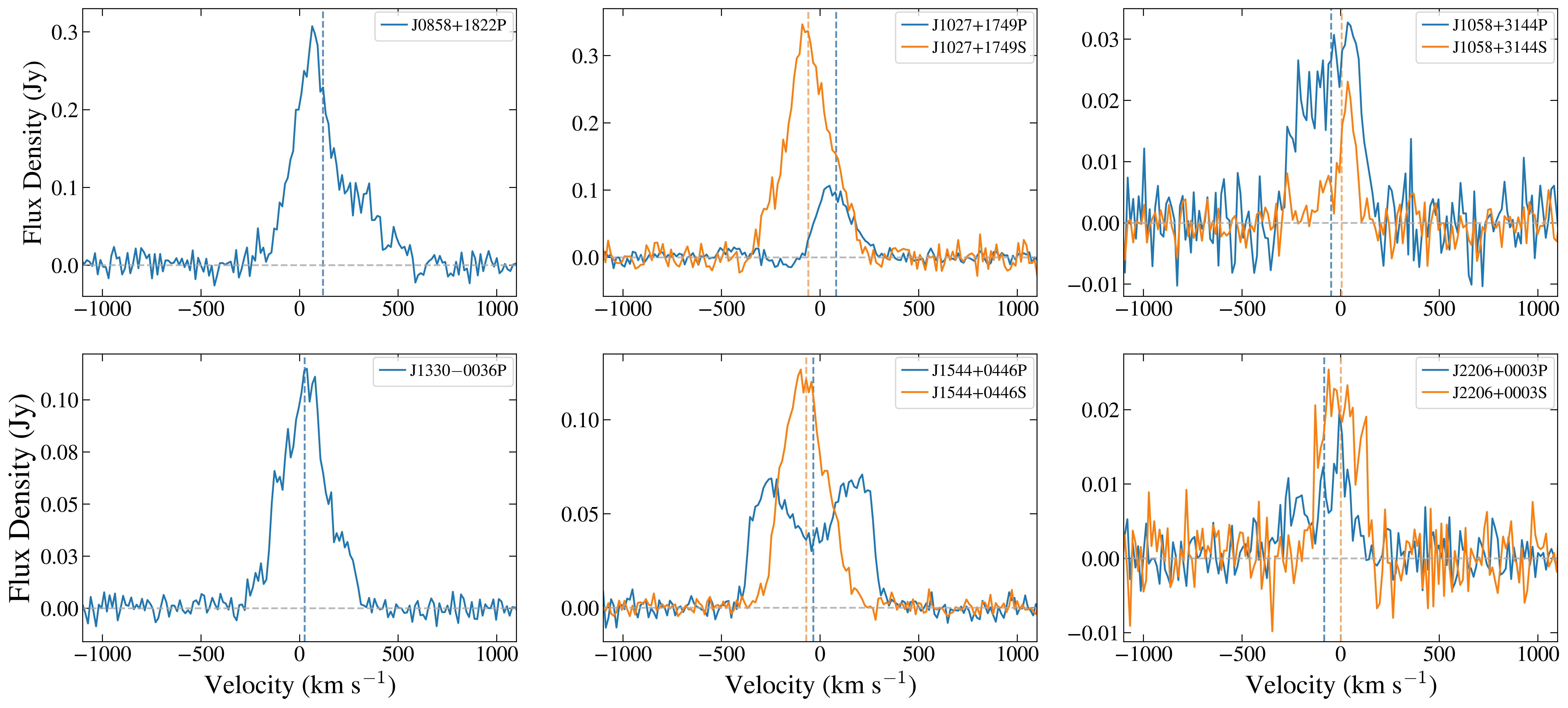}
\caption{CO(2-1) line profile of the detected nuclei, blue for the primary and orange for the secondary (when detected).
The vertical dashed lines mark the intensity-weighted mean velocity. 
\label{fig:profile}
}
\end{figure*}

\section{Detection of Circumnulear Molecular Gas} \label{sec:analysis}
\subsection{Molecular Gas Measurement} \label{subsec:det}
We exported the data to a CASA-compatible format and performed CO measurement in CASA (version 6.3.0.48, \citealp{McMullin2007}). We removed all the pixels in the datacube with S/N less than 3 and created moment maps for each target using the {\it immoments} task. We further used Source Finding Application (SoFiA, \citealp{Serra2015}) to highlight the source region and generated a new version of the maps for better visualization. 
The resultant line intensity (moment 0), velocity (moment 1), and velocity dispersion (moment 2) maps are shown in Figure~\ref{fig:imagepair}. 
Ten of the 14 nuclei have a clear detection of CO(2-1) emission. Specifically, four dual-AGN systems (J1027+1749, J1058+3114, J1544+0446, and J2206+0003) have both nuclei detected, two systems (J0858+1822 and J1330$-$0036) have only one nucleus detected, and the remaining system (J0022+0022) has no CO detection in either nucleus, which is not shown in Figure~\ref{fig:imagepair}.
The CO emission of most detected nuclei are spatially compact. 
Figure~\ref{fig:profile} shows the CO(2-1) line profiles of the detected nuclei, which are constructed by summing up pixels above three times the RMS level in the moment 0 map.
Below we provide brief remarks on the individual pairs, and we defer a detailed study of the gas spatial distribution and kinematics to a future work. 

For J0858+1822, molecular gas is only detected in the primary nucleus, but there is an appreciable offset ($\sim 2\arcsec$) between the optical position of primary nucleus and the peak of CO emission in the intensity map.
The peak of the velocity dispersion map is also significantly offset from the optical nucleus. 
The velocity field appears to be dominated by a rotating structure (i.e., the redshifted and blueshifted components are roughly symmetrical), but the line profile exhibits an excess on the redshifted (positive) side up to $\sim 400\rm~km~s^{-1}$ from the velocity peak. Such non-circular motions might be due to an infalling stream or an outflow of molecular gas.
No clear excess can be associated with the secondary nucleus, although this might be partially owing to the close separation between the two nuclei. 
Overall these features suggest a merger system in the late stage, where the molecular gas in the individual galaxies are being mixed but not yet fully settled. 
For J1027+1749, molecular gas is detected in both nuclei but is more concentrated around the secondary nucleus. The velocity field around each nucleus is consistent with a rotating structure (e.g., a disk or a tori) but also shows faint non-circular features, which are also evident in the line profile. 
For J1058+3114, which has the best angular resolution among all pairs, molecular gas is detected in both nuclei. Additionally, there is a clump of molecular gas in the north of the primary nucleus, which is spatially coincident with the northern spiral arm of the primary galaxy and has a velocity offset ($\sim 300$ km s$^{-1}$) from the two nuclei. The velocity field around the primary nucleus is consistent with a resolved rotating structure, but the rather flat-topped line profile suggests that molecular gas is not centrally concentrated.   
The secondary nucleus is rather compact, but also exhibits small deviations from an otherwise narrow and symmetric line profile.  
For J1330$-$0036, molecular gas is only detected in the primary nucleus. 
Both the morphology and line profile suggest the presence of extended features in addition to the bright compact core, which are mostly seen between the two nuclei. This may indicate an ongoing mixing of gas from the two galaxies.
A faint patch of CO emission can also be seen around the secondary nucleus in the intensity map, but it is possibly contaminated by an interference fringe, thus we conservatively treat it as a non-detection.
For J1544+0446, molecular gas is detected in both nuclei. The primary nucleus exhibits a double-peaked line profile, which is reminiscent of a centrally-depleted rotating disk or a ring. 
The full-width at half-maximum (FWHM) of the line reaches $\sim 600\rm~km~^{-1}$, the highest among all detected nuclei. 
Moreover, the peak of velocity dispersion is offset from the optical nucleus and reaches $\sim 230 {\rm~km~s^{-1}}$, also the highest value seen among all nuclei, which may indicate the presence of tidal shocks or an outflow.
The secondary nucleus shows a centrally peaked and symmetric profile.
For J2206+0003, molecular gas is detected in both nuclei, which appear well separated from each other. 
The velocity field of both nuclei indicates a compact rotating structure, 
and the flat-topped line profile seen in the secondary nucleus suggests that this structure is centrally depleted.

We quantify the CO(2–1) emission from each nucleus using the CASA task {\it IMFIT}. On the intensity map, each nucleus is fitted by a two-dimensional Gaussian model, which is a good approximation for the observed CO intensity distribution. The CO(2–1) luminosity is then calculated following \cite{Solomon2005}, 
\begin{equation}
    L_{\rm CO}^{'}=3.25\times 10^7~S_{\rm CO}~\Delta V~\nu_{\rm obs}^{-2}~D_{\rm L}^{2}~(1+z)^{-3}
\end{equation}
where $L_{\rm CO}^{'}$ is the CO line luminosity in units of ${\rm K~km~s^{-1}~pc^{2}}$, $S_{\rm CO}\Delta V$ is the CO integrated flux density in units of ${\rm Jy~km~s^{-1}}$, $\nu_{\rm obs}$ is the observed frequency in GHz, and $D_{\rm L}$ is the luminosity distance in Mpc.
For the non-detected nuclei, we estimate a $3\sigma$ upper limit of $S_{\rm CO}\Delta V$ by $3\sigma_{\rm ch}(\Delta V_{\rm ch} \times \Delta V_{\rm line})^{1/2}$ \citep{Seaquist1995, Wagg2007}, where the $\sigma_{\rm ch}$ and $\Delta V_{\rm ch}$ are the channel RMS and width, respectively, $\Delta V_{\rm line}$ is the line-of-sight velocity range of the CO line, assuming the same value of the neighboring detected nucleus (for the case of J0022+0022, where both nuclei are non-detected, $\Delta V_{\rm line} = 500\rm~km~s^{-1}$ is assumed).
To derive the equivalent hydrogen column density ($N_{\rm H}$) and the molecular hydrogen mass (${M_{\rm H_2}}$), we adopt the conventional CO-to-H$_2$ conversion factor $X_{\rm CO} = 2 \times 10^{20} {\rm~cm^{-2}~(K~km~s^{-1}})^{-1}$ and $\alpha_{\rm CO} = 4.3~M_{\sun}~({\rm K~km~s^{-1}~pc^{2}})^{-1}$, which are suitable for the CO(1–0) line \citep{Bolatto2013}.
To further convert into CO(2-1), we have assumed an intrinsic CO(2-1)/CO(1-0) intensity ratio $R_{21} = 1$, which is not atypical for galactic nuclei \citep[e.g.,][]{Leroy2009,Li2019}. 
The hence derived beam-averaged CO-based column density, $N_{\rm H} = X_{\rm CO} \times R_{21} \times L_{\rm CO}^{'}/ \Omega$, ranges from $4.6 \times 10^{21}$ to $1.79 \times 10^{23} {\rm~cm^{-2}}$ for the 10 detected nuclei (Table \ref{table:CO}). Here $\Omega$ is the area of the synthesized beam. We note that the actual column density could be higher if the detected CO emission arises from a region significant smaller than the beam (see further discussions in Section~\ref{sec:discuss}).
For the four non-detected nuclei, the $3\sigma$ upper limits of $N_{\rm H}$ is similarly converted from the upper limit of $L_{\rm CO}^{'}$, which ranges from $0.8 \times 10^{21}$ to $2.7 \times 10^{21} {\rm~cm^{-2}}$.
The corresponding values of $M_{\rm H_2}$, ranging between $(0.07-20.75) \times 10^9\rm~M_{\odot}$, are also listed in Table \ref{table:CO}.
We note that the uncertainty related to $X_{\rm CO}$ and $R_{21}$, which should not exceed a factor of a few, does not affect the main conclusion drawn below.

\begin{deluxetable*}{lccccccccc}
\tabletypesize{\footnotesize}
\tablecaption{NOEMA CO Observations of Seven Dual-AGNs \label{table:CO}}
\tablehead{
\colhead{Name} & 
\colhead{Date} & 
\colhead{Exp} & 
\colhead{Freq} & 
\colhead{Beam size} & 
\colhead{PA} & 
\colhead{$V_{\rm mean}$} & 
\colhead{$S_{CO}\Delta V$} & 
\colhead{$N_{\rm H}$} & 
\colhead{${\rm M_{H_2}}$}
}
\colnumbers
\startdata
J0022+0022S & 2020-10-24 &  0.4  & 215.3 & $1.43 \times 0.71$  &    5.4  & \nodata &  $<$ 0.88          &  $<$ 0.27          &  $<$ 0.23          \\
J0022+0022P & 2020-10-24 &  0.4  & 215.3 & $1.43 \times 0.71$  &    5.4  & \nodata &  $<$ 0.81          &  $<$ 0.25          &  $<$ 0.21          \\
\hline
J0858+1822S & 2020-11-01 &  0.6  & 217.7 & $3.63 \times 1.38$  &   169.1 & \nodata &  $<$ 1.44          &  $<$ 0.09          &  $<$ 0.26          \\
J0858+1822P & 2020-11-01 &  0.6  & 217.7 & $3.63 \times 1.38$  &   169.1 &  118.9  &  $ 74.30 \pm 1.20$ &  $ 4.64 \pm 0.07$  &  $ 13.25 \pm 0.21 $ \\
\hline
J1027+1749P & 2020-11-23 &  0.9  & 216.1 & $4.61 \times 0.95$  &    4.2  &   81.2  &  $ 18.23 \pm 0.98$ &  $ 1.29 \pm 0.07$  &  $ 4.15 \pm 0.22 $ \\
J1027+1749S & 2020-11-23 &  0.9  & 216.1 & $4.61 \times 0.95$  &    4.2  & $-$59.9 &  $ 90.60 \pm 1.50$ &  $ 6.49 \pm 0.11$  &  $ 20.68 \pm 0.34$ \\
\hline
J1058+3144S & 2020-11-05 &  1.1  & 214.9 & $1.07 \times 0.75$  &   40.6  &   5.2   &  $  2.18 \pm 0.08$ &  $ 1.03 \pm 0.04$  &  $ 0.60 \pm 0.02 $ \\
J1058+3144P & 2020-11-05 &  1.1  & 214.9 & $1.07 \times 0.75$  &   40.6  & $-$48.3 &  $  5.83 \pm 0.08$ &  $ 2.97 \pm 0.04$  &  $ 1.58 \pm 0.02 $ \\
\hline
J1330-0036S & 2020-11-22 &  2.6  & 218.7 & $2.29 \times 0.80$  &    8.4  & \nodata &  $<$ 0.45          &  $<$ 0.08          &  $<$ 0.07          \\
J1330-0036P & 2020-11-22 &  2.6  & 218.7 & $2.29 \times 0.80$  &    8.4  &   26.3  &  $ 25.61 \pm 0.19$ &  $17.84 \pm 0.13$  &  $ 3.85 \pm 0.03$ \\
\hline
J1544+0446S & 2021-01-06 &  1.0  & 221.2 & $3.51 \times 0.89$  &    8.1  & $-$70.6 &   $ 34.30 \pm 0.34$ & $ 3.49 \pm 0.03$  &  $ 3.07 \pm 0.03 $ \\
J1544+0446P & 2021-01-06 &  1.0  & 221.2 & $3.51 \times 0.89$  &    8.1  & $-$34.6 &  $ 40.80 \pm 0.55$ &  $12.62 \pm 0.17$  &  $ 3.59 \pm 0.05$ \\
\hline
J2206+0003S & 2020-10-24 &  0.3  & 220.3 & $1.52 \times 0.94$  &    0.0  &   0.7   &  $  4.11 \pm 0.03$ &  $ 0.91 \pm 0.01$  &  $  0.46 \pm 0.01 $ \\
J2206+0003P & 2020-10-24 &  0.3  & 220.3 & $1.52 \times 0.94$  &    0.0  & $-$83.8 &  $  2.06 \pm 0.03$ &  $ 0.46 \pm 0.01$  &  $  0.22 \pm 0.01 $ \\
\enddata
\tablecomments{(1) Abbreviated names. `P' and `S' denote the primary and secondary nuclei according to the stellar velocity dispersion; (2) observation date; (3) effective on-source time based on visibilities after flagging data, in units of hours; 
(4) observed frequency, in unit of GHz;
(5)synthesized beam size, in units of $\arcsec \times \arcsec$; (6) position angle of the synthesized beam, in units of degrees; 
(7) intensity-weighted mean velocity, relative to the tuning frequency, in unit of km ${\rm s^{-1}}$;
(8) NOEMA CO(2-1) 230 GHz flux density, in units of Jy$\rm~km~s^{-1}$; for non-detected nuclei, 3$\sigma$ upper limits are given; 
(9) hydrogen column density converted from CO(2-1), in units of $10^{22} {\rm~cm^{-2}}$; 
(10) derived molecular hydrogen mass, in units of $10^{9} {\rm~M_\odot}$.
}
\end{deluxetable*}

\subsection{Correlations with X-ray and Optical Properties} \label{subsec:flux}
\begin{figure*}
\centering
\includegraphics[width=0.95\textwidth]{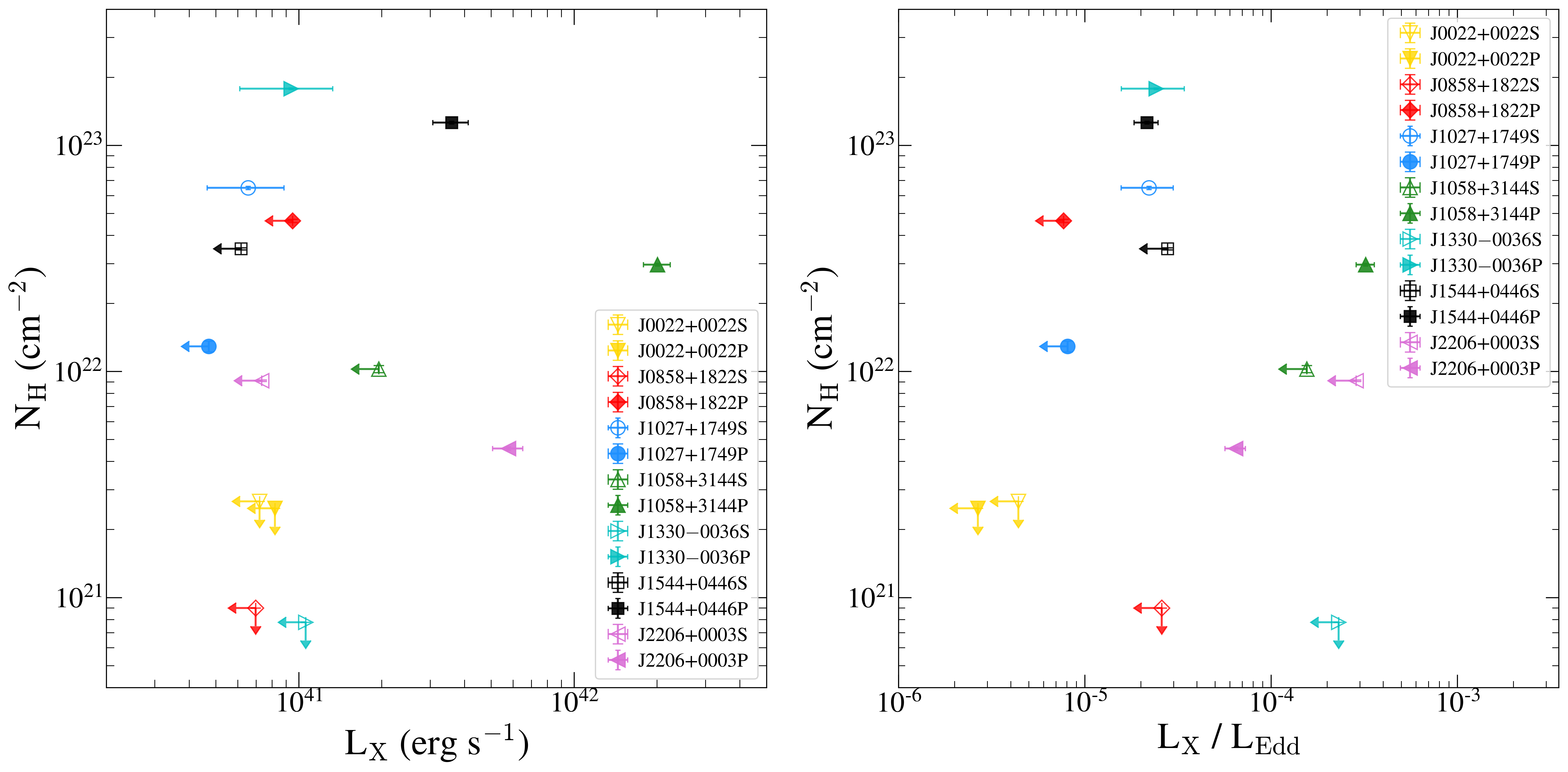}  
\caption{Beam-averaged CO-based equivalent hydrogen column density vs. 2--10 keV X-ray luminosity ({\it left panel}) and X-ray Eddington ratio ({\it right panel}) for the 14 nuclei in 7 pairs of dual-AGN. 3$\sigma$ upper limits of the non-detected (in CO or X-rays) nuclei are denoted by arrows. The primary and secondary nuclei are marked by solid and open symbols, respectively.
}
\label{fig:nhlx}
\end{figure*}

One of the primary goals of this work is to understand whether the apparently low X-ray luminosities of the dual-AGNs are due to strong circumnuclear obscuration. 
Figure \ref{fig:nhlx} displays the equivalent hydrogen column density derived from the CO(2-1) emission versus the 2--10 keV X-ray luminosity ($L_{\rm X}$, left panel) and the X-ray Eddington ratio ($f_{\rm Edd,X} = L_{\rm X}/L_{\rm Edd}$, right panel) for each nucleus in our sample.
The Eddington ratio is defined as $L_{\rm X}$ normalized by the Eddington luminosity ($L_{\rm Edd}$) of the given nucleus, originally calculated by \citetalias{Hou2020} according to the black hole mass-stellar velocity dispersion relation and here updated for the revised stellar velocity dispersion.
The 14 nuclei have very moderate $L_{\rm X} \lesssim 2\times10^{42} {\rm~erg~s^{-1}}$ and highly sub-Eddington $f_{\rm Edd,X} \lesssim 3\times10^{-4}$. 
We note that the X-ray luminosity, or its upper limit, was primarily estimated by assuming a canonical spectral model, i.e., an absorbed power-law with a presumed absorption column density $N_{\rm H,X} = 10^{22}\rm~cm^{-2}$ and a photon-index of 1.7 \citepalias{Hou2020}. Such a treatment was appropriate for the limited number of X-ray counts for most nuclei, but could have significantly underestimated the intrinsic absorption column density.

Nevertheless, it can be seen from Figure \ref{fig:nhlx} that the CO-based $N_{\rm H}$ has typical values of $\sim10^{22} {\rm~cm^{-2}}$ and in all cases below $\sim2\times10^{23}{\rm~cm^{-2}}$. 
Moreover, four of the hard X-ray-undetected nuclei, J0022+0022S, J0022+0022P, J0858+1822S and J1330-0036S, are also undetected in CO(2-1), and the two nuclei with the highest $L_{\rm X}$, J1058+3144P and J2206+0003S, have only a moderate CO-based $N_{\rm H}$. 
In the two nuclei with $N_{\rm H} \sim (1-2) \times 10^{23} {\rm~cm^{-2}}$, there might be substantial intrinsic X-ray obscuration, but raising the assumed $N_{\rm H,X}$ from $10^{22} {\rm~cm^{-2}}$ to $10^{23} {\rm~cm^{-2}}$ 
would lead to an increase of $L_{\rm X}$ and $f_{\rm Edd,X}$ by only a factor of $\sim2$.
Overall, the amount of circumnuclear molecular gas is moderate in the present sample of dual-AGNs, which argues against strong X-ray obscuration being prevalent in these systems (see further discussions in Section~\ref{sec:discuss}).


Lastly, we note the interesting trend that in four pairs the higher $N_{\rm H}$ and $M_{\rm H_2}$ are found with the primary nucleus (marked by solid symbols in Figure~\ref{fig:nhlx}). The exceptions are J1027+1749 and J2206+0003, in which the secondary has an $N_{\rm H}$ about 5 and 2 times higher than that of the primary, respectively. The remaining pair, J0022+0022, has no significant CO detected in either host galaxy. 
Such a trend, if further supported by a larger sample of close pairs, might be understood as the more massive nucleus/host galaxy being more likely to accumulate a substantial amount of cold gas due to a stronger gravitational potential.

\section{Summary and Discussion} \label{sec:discuss}

Utilizing NOEMA CO(2-1) observations, we have conducted a pilot survey of circumnuclear molecular gas in seven dual-AGN systems at a redshift $\sim 0.06$. 
These systems have the smallest projected separation ($r_{\rm p} \lesssim 5$ kpc) among the optically and X-ray selected AGN pairs studied by \citetalias{Hou2020}, providing an important closeup view of AGN activity at or close to the late stage of galaxy mergers.
The high-resolution NOEMA observations are able to resolve the two closely separated nuclei in a pair. Significant CO(2-1) emission is detected in 10 of the 14 nuclei, which traces circumnuclear molecular gas with a sensitivity equivalent to a limiting hydrogen column density of $\sim1.0\times10^{21}{\rm~cm^{-2}}$, assuming a standard CO-to-H$_2$ conversion factor and a CO(2-1)/CO(1-0) intensity ratio of unity. 
The derived equivalent hydrogen column density ranges between $5\times10^{21} - 2\times10^{23}\rm~cm^{-2}$, but mostly at values of a few times $10^{22}\rm~cm^{-2}$.
Most CO-detected nuclei show a compact morphology, but extended CO emission is also found in at least two cases (J1058+3144P and J2206+0003S). 
The CO velocity field and line profile of most nuclei are  consistent with a rotating structure, but non-circular motions are also evident in some cases. 

\citetalias{Hou2020} found that all 14 nuclei have a moderate 2--10 keV luminosity $\lesssim 2\times10^{42}\rm~erg~s^{-1}$ and a low Eddington ratio $\lesssim3\times10^{-4}$ (Figure~\ref{fig:nhlx}). 
This is rather surprising, as galaxy pairs at such small projected separations ($\lesssim 5$ kpc) are expected to have experienced strong gravitational torques, which induce gas inflows to the galactic nucleus and subsequently trigger active SMBH accretion. 
This has been extensively demonstrated by numerical simulations of idealized galaxy mergers \citep[e.g.,][]{Capelo2015,Capelo2017,Solanes2019,Yang2019}, although until recently numerical simulations still rely on prescriptions of sub-grid physics to account for accretion, radiation and feedback of the SMBH \citep{DeRosa2019}. 
One possibility to reconcile the apparently low X-ray luminosities is strong absorption typical of Compton-thick AGNs, which have a line-of-sight column density $N_{\rm CT} \gtrsim 10^{24}\rm~cm^{-2}$, such that X-ray photons with an energy below a few keV are completely obscured. 
However, none of the 14 nuclei show a CO-based hydrogen column density above the Compton-thick threshold, and in fact, the majority fall bellow $10^{23}\rm~cm^{-2}$.
A potential caveat is that this $N_{\rm H}$ is derived by averaging over the synthesized beam of the NOEMA observation, which corresponds to a physical scale of $\lesssim$ few kpc, whereas direct absorption of the X-rays may arise from a smaller (10--100 pc) region around the SMBH.
If the detected CO emission arises from this more compact region, the actual column density could be much greater.
For instance, a recent study by \citet{Garcia-Burillo2021} based on high-resolution Atacama Large Millimeter Array (ALMA) survey of nearby Seyfert galaxies detected spatially resolved dusty molecular tori located around the AGN, 
which have a median torus diameter of $\sim 42$ pc. 
The molecular gas column density, $N({\rm H_2})$, derived from the CO emission, 
shows a typical range ($\sim 8\times 10^{21} - 4\times 10^{23} \rm~cm^{-2}$) similar to our sample, 
but for five low-luminosity AGNs in their sample
($L_{2-10 \rm~keV} \sim 10^{40} \rm~erg~s^{-1}$), the molecular gas column densities are much higher ($\sim 10^{24} \rm~cm^{-2}$).

However, such a case is unlikely relevant to the majority of our sample, for the following considerations. 
Firstly, among the whole sample, the three systems (J0022+0022, J1058+3144 and J2206+0003) with the best physical resolution ($\sim 1.5-2$ kpc) are also among the ones with the lowest CO-based $N_{\rm H}$ (perhaps except for J1058+3144P).
Secondly, among the whole sample, three nuclei have a sufficient number of detected X-ray counts for a spectral analysis. Following \cite{Hou2019}, we adopt an absorbed power-law model to fit the spectra of these nuclei. The best-fit absorption column density ($N_{\rm H,X}$) is found to be $6.4^{+6.3}_{-4.5} \times 10^{22}{\rm~cm^{-2}}$, $3.9^{+3.2}_{-2.7} \times 10^{22}{\rm~cm^{-2}}$, and $5.7^{+23.7}_{-3.7} \times 10^{22}{\rm~cm^{-2}}$, for J1058+3144P, J1544+0446P, and J2206+0003S, respectively.
These values are consistent with the CO-based $N_{\rm H}$ within a factor of $\sim2$, indicating that the latter is a good proxy to the actual absorption column density.
In fact, the CO-based column density and X-ray-based column density are also consistent within a factor of $\sim$2 in the Seyfert galaxies studied by \citet[][Fig. 17 therein]{Garcia-Burillo2021}.
Thirdly, the infrared color can provide an independent probe of luminous AGNs ($L_{\rm X} \gtrsim 10^{43}\rm~erg~s^{-1}$; \citealp{Jarrett2011,Stern2012}), in the standard picture in which a dusty torus absorbs the central engine's intense X-ray/ultraviolet radiation and subsequently re-radiates copious infrared emission.
Specifically, a WISE \citep{Wright2010} color cut, $W1$(3.4 $\mu$m)$-W2$(4.6 $\mu$m) $> 0.8$, robustly separates luminous AGNs from star-forming galaxies \citep{Stern2012}.
It turns out all seven dual-AGNs studied here have $W1–W2 < 0.8$ (Table~\ref{table:info}), with a median value of 0.2. 
Even with the more conservative cut of $W1-W2 > 0.5$ suggested by \citet{Satyapal2014} (see also discussions by \citealp{Blecha2018}), which tolerates weaker AGNs, only one system (J1330$-$0036) satisfies the threshold with its $W1-W2 \approx 0.6$.
This suggests that no intrinsically luminous AGN exists in the majority, if not all, of the seven systems.
We note that given the relatively large WISE point-spread function (FWHM $\sim$ 6$\arcsec$), the two nuclei in any of the seven pairs are unresolved and thus share the same value. Nevertheless, this should not alter the above conclusion. 

Therefore, we conclude that the moderate X-ray luminosities found in the seven dual-AGNs are unlikely due to circumnuclear obscuration, but should reflect the prevalence of weakly accreting SMBHs in these nuclei. 
This may seem surprising for these merging systems at or close to their late evolutionary stage. \citetalias{Hou2020} proposed a plausible explanation which involves a feeding-and-feedback loop, which we elaborate here. 
In this picture, AGN activity is triggered in one or both of the SMBHs by the pericentric passage of the two host galaxies. The subsequent AGN feedback can halt the gas inflow, expel gas from the central region, and suppress subsequent SMBH accretion.
This feedback is likely in a kinetic mode, given that the accretion rates inferred from the X-ray luminosities of the dual-AGNs are generally highly sub-Eddington (Figure~\ref{fig:nhlx}; \citetalias{Hou2020}). Recently, \citet{Peng2022} reported clear evidence of ongoing AGN feedback in the candidate triple-AGN system, SDSS
J0849+1114, in which kpc-scale, energetic radio jets are present in two of the three optical nuclei. 
The estimated jet energetics (a few $10^{55}$ erg) is comparable to that needed to unbind a cold gas of $\sim10^8\rm~M_{\odot}$ from the central $\sim$100 pc of the host galaxy. 
Thus this fine example is quite relevant to the feedback scenario discussed here.
Alternatively, the AGN feedback might be predominantly radiative, which pushes out the dusty gas via radiative pressure \citep[e.g.,][]{Ricci2022}. However, for the radiative mode to be dominant, the accretion rate should be close to the Eddington limit, which must be rare given the statistical result of \citetalias{Hou2020}.

Two potential issues with the AGN feedback scenario deserve further remarks.
First, not {\it all} 14 nuclei are deficient in circumnuclear molecular gas. Indeed two nuclei (J0858+1822P and J1027+1749S) have $\sim10^{10}\rm~M_{\odot}$ of gas detected within the central few kpc, which is comparable to gas-rich normal galaxies. On the other hand, at least three nuclei (J1058+3144P, J1544+0446S, J2206+0003S) show a flat-topped line profile, which suggests that the gas is not centrally concentrated, consistent with depletion of gas in the close vicinity of the SMBH.
Higher-resolution CO observations, e.g., possibly afforded by ALMA, would be helpful to confirm a paucity of cold gas within the gravitational influence radius ($\lesssim$ 100 pc) of the SMBHs in these nuclei.
Second, for the feedback scenario to be compatible with the low luminosities and low accretion of essentially all nuclei studied here, it would require a relatively short phase of active accretion and strong feedback. 
This is most feasible with the second pericentric passage of the merging pair, which simulations predict to boost AGN activity for a few tens of Myr when the two nuclei are separated by a few kpc  (e.g., \citealp{Capelo2015}).
After that, the accretion rate drops back to and maintains at a low level until the two galaxies finally merge. 
High-resolution radio and optical spectroscopic observations are warranted to search for direct evidence of AGN feedback in these and other kpc-scale dual-AGNs, especially the ones with relatively high Eddington ratios.


\begin{acknowledgments}
This work is based on observations carried out under project number S20BJ with the IRAM NOEMA Interferometer. IRAM is supported by INSU/CNRS (France), MPG (Germany) and IGN (Spain).
Special acknowledgement is dedicated to the late Dr. Yu Gao, whose encouragement made this NOEMA program possible.
M.H. wishes to thank Vinod Arumugam for his assistance with the NOEMA observation and data reduction, Yanmei Chen, Y. Sophia Dai, Fengyuan Liu and Zhiyu Zhang for helpful discussions.
M.H. is supported by the National Natural Science Foundation of China (12203001) and the fellowship of China National Postdoctoral Program for Innovation Talents (grant BX2021016). 
L.C.H. was supported by the National Science Foundation of China (11721303, 11991052, 12011540375) and the China Manned Space Project (CMS-CSST-2021-A04, CMS-CSST-2021-A06). 
Z.L. acknowledges support by the National Key Research and Development Program of China (2017YFA0402703) and National Natural Science Foundation of China (11873028). X.L. acknowledges support from NSF grants AST-2108162 and AST-2206499.
\end{acknowledgments}


\bibliography{NOEMA}
\bibliographystyle{aasjournal}

\end{document}